\documentclass[runningheads]{llncs}

\usepackage{eccv}

\usepackage{eccvabbrv}

\usepackage{graphicx}
\usepackage{booktabs}

\usepackage{amsmath}
\usepackage{amssymb}
\usepackage{mathtools}

\usepackage{graphicx}

\usepackage{algorithm}
\usepackage{algorithmic}

\usepackage{amsfonts}

\usepackage{amssymb}
\usepackage{array}
\usepackage{subcaption}
\usepackage{caption}
\usepackage{textcomp}
\usepackage{url}
\usepackage{verbatim}
\usepackage{graphicx}
\usepackage[dvipsnames,table]{xcolor}
\usepackage{soul}
\usepackage{enumitem}

\usepackage{arydshln}
\usepackage{fancyhdr}
\usepackage{varwidth}
\usepackage{arydshln}
\usepackage{multirow}
\usepackage{mathtools}
\usepackage{algorithmic}
\usepackage{algorithm}

\usepackage{datetime2}
\usepackage{pifont}
\usepackage{tabularx}

\usepackage[accsupp]{axessibility}  

\usepackage{newfloat}
\usepackage{listings}

\usepackage{kotex}
\usepackage{soul}

\usepackage{tikz}

\newcommand{\sys}{\textsf{\small SPOILER}}

\definecolor{dg}{rgb}{0.0, 0.65, 0.0}
\newcommand{\highlight}[1]{\textcolor{dg}{\textbf{#1}}}

\newcommand{\remove}[1]{}

\usepackage{datetime}
\newdateformat{monthdayyeardate}{%
  \monthname[\THEMONTH] \THEDAY, \THEYEAR}

\usepackage{booktabs}

\usepackage{hyperref}

\usepackage{orcidlink}

\begin{document}

\title{SPOILER: TEE-Shielded DNN Partitioning of On-Device Secure Inference with Poison Learning} 

\titlerunning{Abbreviated paper title}

\author{Donghwa Kang\inst{1} \and Hojun Choe\inst{1} \and Doohyun Kim\inst{1} \and Hyeongboo Baek\inst{2} \and Brent ByungHoon Kang\inst{1}}

\authorrunning{D. Kang et al.}

\institute{Korea Advanced Institute of Science and Technology (KAIST), Republic of Korea\and
University of Seoul, Republic of Korea
}

\maketitle

\begin{abstract}
Deploying deep neural networks (DNNs) on edge devices exposes valuable intellectual property to model-stealing attacks. 
While TEE-shielded DNN partitioning (TSDP) mitigates this by isolating sensitive computations, existing paradigms fail to simultaneously satisfy privacy and efficiency. 
The \textit{training-before-partition} paradigm suffers from intrinsic privacy leakage, whereas the \textit{partition-before-training} paradigm incurs severe latency due to structural dependencies that hinder parallel execution. 
To overcome these limitations, we propose \sys{}, a novel \textit{search-before-training} framework that fundamentally decouples the TEE sub-network from the backbone via hardware-aware neural architecture search (NAS). 
\sys{} identifies a lightweight TEE architecture strictly optimized for hardware constraints, maximizing parallel efficiency. 
Furthermore, we introduce \textit{self-poisoning learning} to enforce logical isolation, rendering the exposed backbone functionally incoherent without the TEE component. 
Extensive experiments on CNNs and Transformers demonstrate that \sys{} achieves state-of-the-art trade-offs between security, latency, and accuracy.
\end{abstract}

\section{Introduction}
\label{sec:intro}

The proliferation of deep neural networks (DNNs) has made on-device AI—from smartphone face recognition to autonomous delivery—ubiquitous~\cite{lecun2015deep}.
Processing inference at the edge offers critical benefits: reduced latency, bandwidth conservation, and data privacy.
However, this decentralized architecture exposes models to ``white-box'' environments where physical access is assumed.
Unlike cloud-based black-box models, on-device models are vulnerable to model stealing (MS) attacks, where adversaries clone the model's functionality, compromising the provider's intellectual property (IP)~\cite{tramer2016stealing, orekondy2019knockoff, juuti2019prada}.

\begin{figure}[t]
    \centering
    \includegraphics[width=0.8\linewidth]{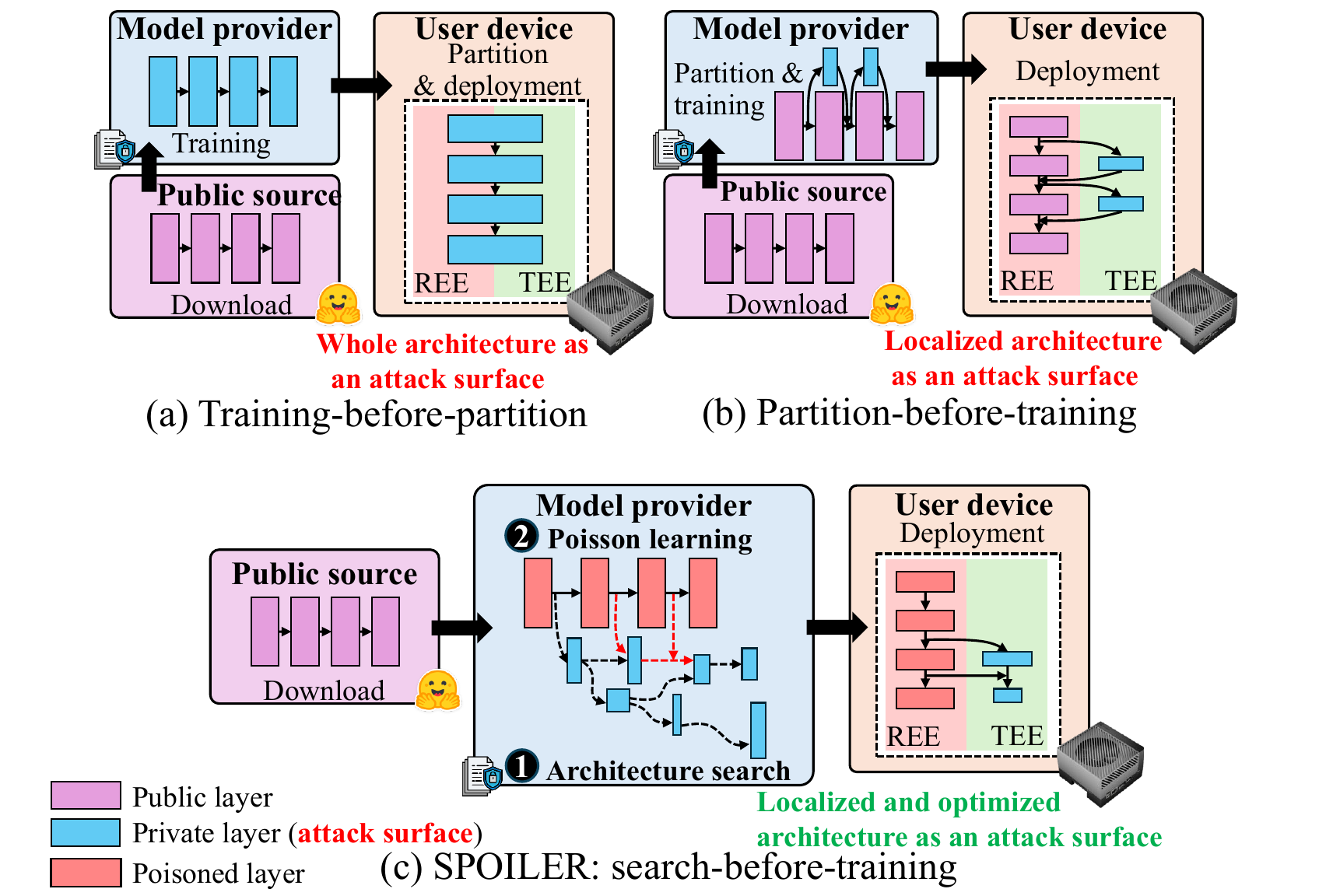}
\caption{The evolution of TSDP paradigms. 
(a) TBP transforms the entire open-source model (purple) into private parameters (blue) through full training, exposing the whole architecture as an attack surface. 
(b) PBT localizes the attack surface by training only the TEE-isolated layers (blue) while retaining the public backbone (purple), yet it still fails to address (model) structural dependencies and hardware heterogeneity. 
(c) \sys{} (ours) introduces the SBT paradigm, which actively searches for a hardware-aware architecture (via (\ding{182}) NAS) for TEE and enforces logical isolation for REE through (\ding{183}) self-poisoning learning.} 
    \label{fig:TSDP_paradigm}
\end{figure}

To mitigate these threats, trusted execution environments (TEEs)~\cite{arm_trustzone, costan2016intel} have been adopted to shield model weights.
However, executing the \textit{entire} model within a TEE incurs prohibitive latency overheads (up to $50\times$) compared to GPU-accelerated execution~\cite{tramer2018slalom, mo2020darknetz}.
To enable practical TEE-based secure inference, a framework must satisfy four imperative requirements:
(i) \textbf{Security}: preventing adversaries from replicating the private model's functionality using components exposed in the rich execution environments (REE);
(ii) \textbf{Efficiency}: minimizing latency while maintaining accuracy comparable to REE-only execution;
(iii) \textbf{Feasibility}: operating within the strict hardware constraints of commodity TEEs (e.g., secure memory capacity); and
(iv) \textbf{Generality}: ensuring applicability across diverse architectures (e.g., CNNs, Transformers) without model-specific dependencies.

Existing TEE-shielded DNN partitioning (TSDP) paradigms fail to simultaneously satisfy these goals.
The training-before-partition (TBP) paradigm (Figure~\ref{fig:TSDP_paradigm}(a)) splits a pre-trained model but fails security, as the REE-offloaded layers retain residual private information; remedying this via heavy obfuscation~\cite{tramer2018slalom, lin2020chaotic} violates efficiency.
Conversely, the partition-before-training (PBT) paradigm (Figure~\ref{fig:TSDP_paradigm}(b)) advances security by isolating TEE components pre-training, yet it remains critically flawed by an inherent \textit{structural dependency}.
Even when employing sophisticated compression techniques like knowledge transfer, these methods typically construct the TEE sub-network by mimicking the backbone's topology (e.g., distilling GPU-optimized blocks into CPU-bound TEEs).
This inescapable architecture-hardware mismatch dictates inefficient execution flows—such as bidirectional slicing~\cite{li2022teeslice} or lock-step dual branches~\cite{xiao2023tbnet}—which cause severe synchronization bottlenecks and GPU idling.
Ultimately, PBT's inability to break free from the backbone's structural legacy imposes a hard ceiling on efficiency, necessitating a paradigm shift.

To transcend these limitations, we propose \sys{}, which exploits a new paradigm: search-before-training (SBT) (Figure~\ref{fig:TSDP_paradigm}(c)).
Unlike prior approaches bound by the fixed topology of the backbone, \sys{} fundamentally \textit{decouples} the TEE sub-network using hardware-aware neural architecture search (NAS)~\cite{zoph2016neural, cai2018proxylessnas}.
This allows us to discover extremely lightweight, hardware-optimized layers that drastically reduce latency.
However, the limited capacity of such compact sub-networks poses a challenge to maintaining high accuracy, necessitating collaborative training with the REE backbone.
While this collaboration typically risks exposing private features, we resolve this dilemma via \textit{self-poisoning learning}.
This mechanism creates a symbiotic synergy: NAS achieves extreme efficiency, while poisoning ensures that the requisite collaborative training reinforces logical isolation, rendering the REE backbone functionally incoherent to adversaries without the TEE component.

\begin{itemize}
    \item We identify the fundamental limitations of existing partitioning paradigms, specifically the structural dependency and hardware agnosticism of PBT methods.
    \item We propose the SBT paradigm and the \sys{} framework, utilizing multi-objective NAS to discover TEE-shielded architectures optimized for specific hardware constraints.
    \item We introduce a self-poisoning learning strategy that neutralizes model stealing attacks by rendering the exposed REE components functionally incoherent.
    \item Extensive evaluations demonstrate that \sys{} achieves state-of-the-art (SOTA) trade-offs between latency, accuracy, and security compared to existing TSDP solutions.
\end{itemize}

\begin{figure}[t]
    \centering
    \includegraphics[width=0.9\linewidth]{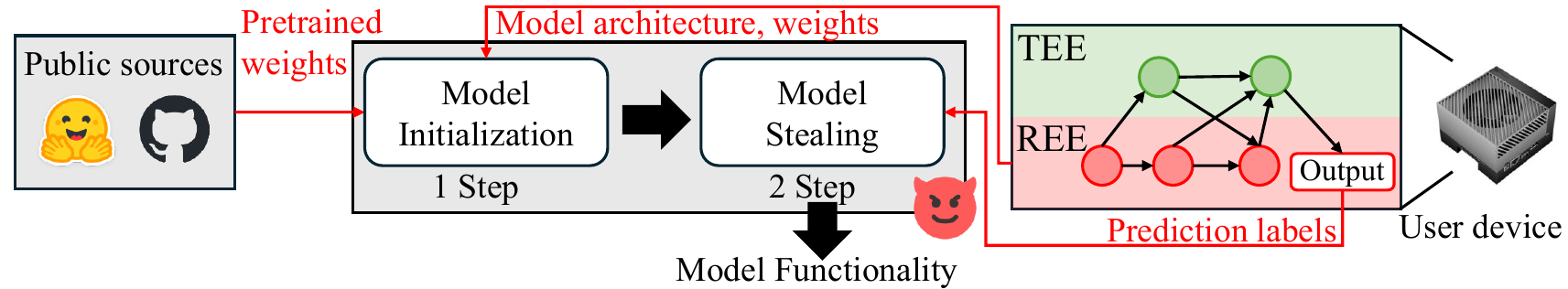}
    \caption{Illustration of the two-step model stealing pipeline. 
In Step 1 (model initialization), the adversary initializes a shadow model by combining a public pre-trained model with parameters exposed in the REE. 
In Step 2 (model stealing), the adversary trains a surrogate model via knowledge distillation using query responses from the victim model, effectively cloning its functionality.}
\label{fig:threat}
\end{figure}

\begin{figure*}[t]
    \centering
    \includegraphics[width=0.93\linewidth]{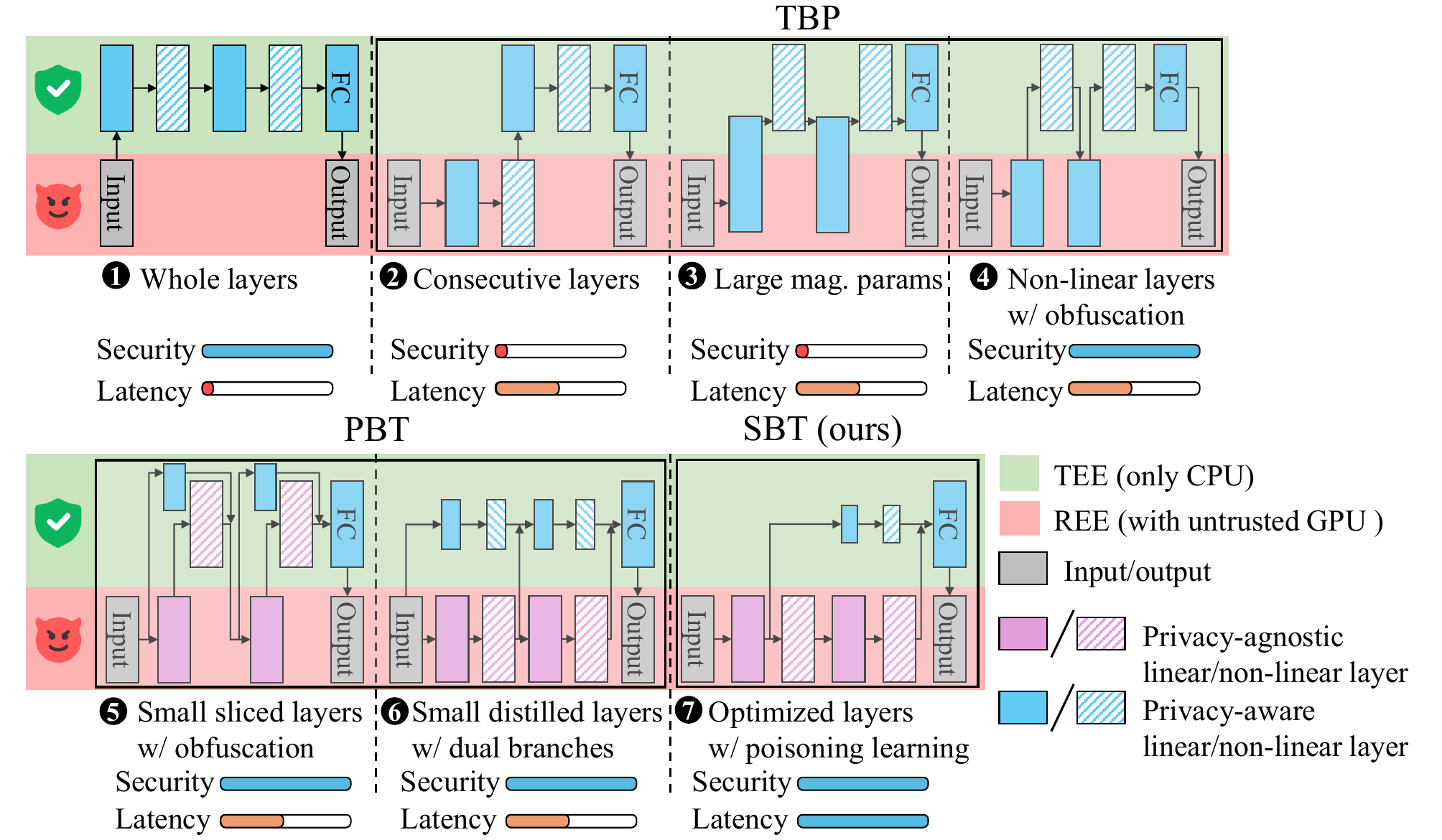}
    \caption{Taxonomy of TSDP methodologies. 
Executing the whole model in the TEE (\ding{182}) incurs prohibitive latency. 
TBP methods partition layers (\ding{183}) or weights (\ding{184}) after training. 
To prevent leakage, some apply obfuscation to REE weights and inputs (\ding{185}). 
PBT methods isolate components pre-training via slicing (\ding{186}) or dual branches (\ding{187}). 
However, these fail due to leakage, overhead, or rigidity. 
In contrast, \sys{} employs hardware-aware NAS and self-poisoning to optimize security and efficiency (\ding{188}).}
\label{fig:taxonomy}
\end{figure*}

\section{Preliminaries and Threat Model}
\label{sec:back_threat}

\subsubsection{TEE.}
TEE is a hardware-isolated enclave designed to guarantee the confidentiality and integrity of code and data~\cite{sabt2015trusted, costan2016intel}.
Operating independently from REE, it communicates solely via secure monitor calls (SMC)~\cite{smc_convention}, ensuring that sensitive assets remain protected from privileged REE adversaries~\cite{pinto2019demystifying}.
ARM TrustZone~\cite{arm_trustzone} is a representative TEE implementation widely deployed in commodity edge devices.

\subsubsection{Defender.}
We assume the defender is a model provider aiming to protect the private model's functionality from unauthorized access~\cite{orekondy2019knockoff, juuti2019prada}.
Our objective is to mitigate white-box attacks on deployed models, effectively downgrading the threat level to a black-box setting where the TEE exposes only final prediction labels.
We assume the victim model utilizes a public pre-trained backbone to enhance accuracy.
The private model is securely provisioned to the user device, after which the TEE selectively offloads non-sensitive computations to the REE (GPU)~\cite{mo2020darknetz, sun2023shadownet}.
Consistent with prior works~\cite{xiao2023tbnet, mo2021serdab}, we exclude user input privacy and physical side-channel attacks from our scope.

\subsubsection{Attacker and attack scenario.}
As illustrated in Figure~\ref{fig:threat}, we consider an adversary with root privileges in the REE who employs a two-step pipeline for MS~\cite{jagielski2020high, tramer2016stealing}.
First, in model initialization (Step 1), the adversary constructs a shadow model by combining a public model (sharing the victim's backbone architecture) with the weights exposed in the REE.
Next, in model stealing (Step 2), the adversary trains a surrogate model via knowledge distillation, querying the victim model with an auxiliary dataset to generate input-prediction pairs for fine-tuning.

\section{Motivation and Design Principles}
\label{sec:motivation}

\begin{figure}[t]
    \centering
    \includegraphics[width=0.9\linewidth]{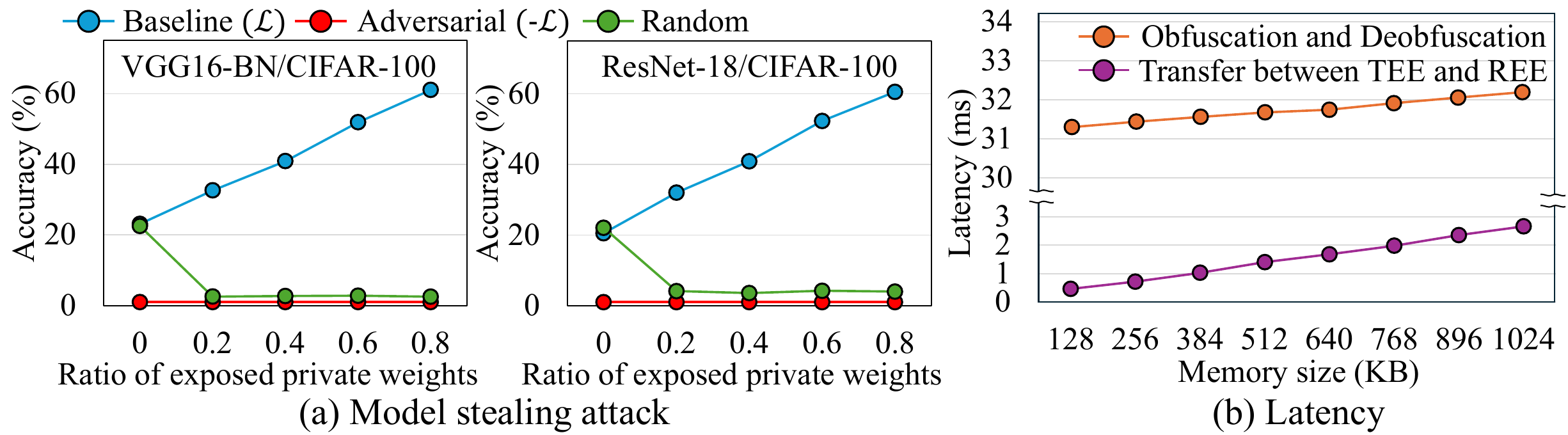}
    \caption{Motivational analysis on NVIDIA Jetson Orin. (a) Accuracy of surrogate models under three settings: targeting a model trained with $\mathcal{L}$ (blue) and $-\mathcal{L}$ (red), and a randomly initialized model (green). Notably, as the ratio of private weights exposed in the REE rises, the accuracy of the baseline surrogate (blue) increases, indicating intrinsic leakage in TBP methods. In contrast, the adversarial and random models yield low accuracy due to a lack of task-aware information.
    (b) The latency of cryptographic obfuscation significantly exceeds that of TEE-REE data transfer; however, the data transfer cost itself remains a non-negligible latency bottleneck.}
    \label{fig:motiv}
\end{figure}

Figure~\ref{fig:taxonomy} categorizes existing TSDP approaches into paradigms based on their defense strategies.
While executing the entire model within the TEE (\ding{182}) theoretically guarantees perfect security, it incurs prohibitive latency due to the lack of GPU acceleration, fundamentally violating efficiency~\cite{tramer2018slalom, mo2020darknetz}.
Consequently, partitioning-based approaches have emerged, yet they face critical trade-offs.

\subsubsection{Limitations of TBP: security vs. efficiency.}
TBP paradigm splits a pre-trained model.
Methods that isolate specific layers (\ding{183})~\cite{mo2020darknetz, mo2021serdab} or weights (\ding{184})~\cite{chen2022magnitude} suffer from intrinsic leakage (violation of security).
As shown in Figure~\ref{fig:motiv}(a), training the model to retain semantic features (blue line) increases the surrogate model's accuracy as more of the model is exposed to the REE. In contrast, adversarial training or random (red or green line), which lacks semantic features, degrades the surrogate model's accuracy even below that of a pretrained model.
To mitigate this, approaches like GroupCover~\cite{kim2023groupcover} (or similar obfuscation methods~\cite{tramer2018slalom}) (\ding{185}) employ heavy obfuscation.
However, our measurements on the NVIDIA Jetson Orin~\cite{Orin} reveal that the cryptographic overhead often outweighs data transmission costs (Figure~\ref{fig:motiv}(b)), severely compromising efficiency.
\textbf{P1}: To ensure security and efficiency, the REE must not contain any private layers capable of learning semantic features, thereby eliminating the need for costly obfuscation.

\subsubsection{Limitations of PBT: sequential dependency.}
PBT paradigm isolates TEE components pre-training to avoid leakage.
Methods like TEESlice~\cite{li2022teeslice} (\ding{186}) and TBNet~\cite{xiao2023tbnet} (\ding{187}) adopt this approach but fail to maximize efficiency.
\ding{186} suffers from high synchronization overheads due to frequent TEE-REE communication~\cite{sun2023shadownet}.
Similarly, \ding{187} employs a one-way structure but creates a strong sequential dependency where the REE GPU remains idle while waiting for TEE CPU computations.
\textbf{P2}: To ensure efficiency, TEE-to-REE data transfer must be minimized, and TEE and REE computations must be decoupled to enable parallel execution.

\subsubsection{Lack of feasibility and generality.}
Furthermore, all aforementioned methods (\ding{182}--\ding{187}) rely on static, heuristic-based partitioning.
They optimize for theoretical metrics (e.g., FLOPs) rather than real-world latency under varying hardware constraints (e.g., memory limits, power modes), violating feasibility.
Additionally, they are often tailored to specific architectures (e.g., CNNs with BatchNorm), lacking generality~\cite{xiao2023tbnet}.
To satisfy these principles, our proposed \sys{} (\ding{188}) introduces the SBT paradigm.
\textbf{P3}: To ensure feasibility and generality, the partitioning strategy must be automatically optimized via hardware-aware search to adapt to diverse model architectures and device constraints.

\section{Methodology}
\label{sec:method}

\begin{figure*}[t]
    \centering
    \includegraphics[width=\linewidth]{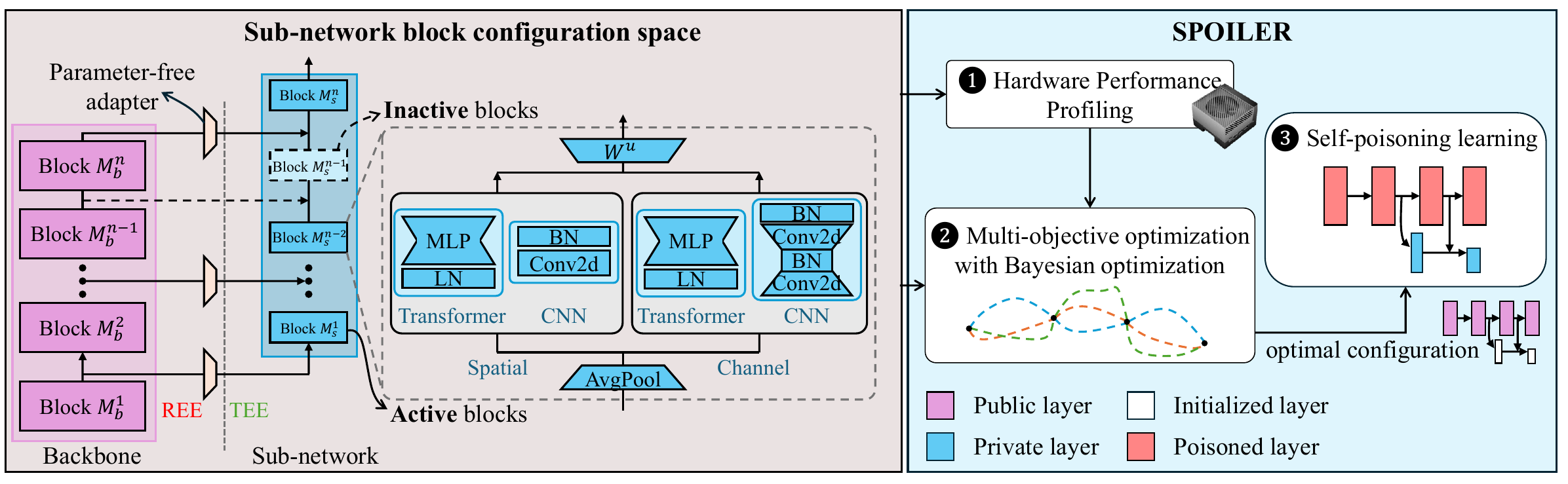}
    \vspace{-0.3cm}
    \caption{Search space and workflow of \sys{}. The search space consists of parameter-free adapters and lightweight blocks designed for TEE constraints.}
    \label{fig:config}
\end{figure*}

In this section, we introduce \sys{}, a novel realization of the SBT paradigm based on P1--P3.
\sys{} operates in two phases:
(1) hardware-aware architecture search, which identifies the optimal TEE sub-network configuration balancing accuracy and latency (Section~\ref{subsec:problem_def}--\ref{subsec:nas}); and
(2) self-poisoning training, which fine-tunes the model to enforce logical isolation, ensuring security without heavy obfuscation (Section~\ref{subsec:poisoning}).

\subsection{Problem Definition}
\label{subsec:problem_def}

The primary objective of \sys{} is to discover a TEE sub-network configuration $a \in \mathcal{A}$ that maximizes accuracy while strictly satisfying hardware constraints.
Formally, we define the bi-objective optimization problem as:

\begin{equation}
  \begin{aligned}
  \max_{a \in \mathcal{A}} \quad & \left( f(a, W^*(a)), -g(a) \right) \\
  \text{s.t.} \quad & W^*(a)=\text{argmin}_{W}\mathcal{L}_{train}(a,W), \;\; \mathcal{H}(a) \leq \mathcal{H}_{limit}
  \end{aligned}
  \label{eq:optimization}
\end{equation}
where $f(\cdot)$ denotes validation accuracy and $g(\cdot)$ represents parallel inference latency (defined in Section~\ref{subsec:design}).
$W^*(a)$ refers to the optimal parameters trained via the loss function $\mathcal{L}_{train}$ (e.g., cross-entropy).
$\mathcal{H}(a)$ encapsulates the hardware resource usage of configuration $a$, which must remain below the hardware constraint threshold $\mathcal{H}_{limit}$ (e.g., available secure memory capacity).

To select the final optimal configuration $a^*$ from the Pareto frontier $\mathcal{A}^*$, we employ a weighted scoring function:
\begin{equation}
    S(a) = \alpha \cdot \tilde{f}(a) + (1-\alpha) \cdot (1-\tilde{g}(a)),
    \label{eq:score}
\end{equation}
where $\tilde{f}$ and $\tilde{g}$ are Min-Max normalized accuracy and latency, and $\alpha$ balances the trade-off.

\subsection{Design of \sys{}}
\label{subsec:design}

\subsubsection{Design principles.}
To adhere to P1, the TEE sub-network is designed as an isolated "side-branch" containing semantic features, ensuring no sensitive parameters reside in the REE.
This decoupling naturally satisfies P2 by enabling parallel execution between the REE backbone (GPU) and the TEE sub-network (CPU).
By enforcing strictly unidirectional data transfer from the REE to the TEE, \sys{} eliminates complex synchronization and obfuscation, maximizing parallel efficiency.
To further address P2, \sys{} minimizes data transfer volume using parameter-free adapters.
Crucially, to maintain P1 compliance, these adapters compress intermediate features via interpolation without introducing learnable parameters in the REE.

\subsubsection{Parallel execution.}
The sub-network can comprise up to the same number of blocks as the backbone (e.g., 12 encoders for ViTs, or pooling stages for CNNs), excluding the final classifier.
Data transmission to the sub-network initiates only after the input passes at least one backbone block, ensuring every transfer is followed by a sub-network block inference.
Let $K$ and $L$ denote the number of unidirectional data transfer points and total backbone blocks (excluding the classifier), respectively.
Let $p_k$ be the backbone block index for the $k$-th data transfer.
The total parallel inference latency for configuration $a$ is formulated as:

\begin{multline}
    g(a) = \sum_{l=1}^{p_1} c^{G}_{l} + c^{A}_{p_1} + \sum_{k=1}^{K-1} \max \left( c^{T}(a_k) + c^{C}(a_k),  \sum_{l=p_k+1}^{p_{k+1}} c^{G}_{l} + c^{A}_{p_{k+1}} \right) \nonumber \\
     + \max \left( c^{T}(a_K) + c^{C}(a_K), \sum_{l=p_K+1}^{L} c^{G}_{l} \right), (K>0),
    \label{eq:latency}
\end{multline}
where $c^{G}_{i}$ and $c^{C}(a_{k})$ denote the execution times of the $i$-th backbone block (GPU) and the $k$-th sub-network block with configuration $a_{k}$ (CPU), respectively.
Additionally, $c^{A}_{i}$ and $c^{T}(a_k)$ represent the adapter execution time for data compression and the REE-to-TEE data transmission time.

\sys{} employs a handshake-based parallel execution mechanism; latency per synchronization interval is dictated by the maximum execution time between the backbone and sub-network.
The total latency $g(a)$ is inherently lower-bounded by the pure backbone execution time, $\sum_{l=1}^{L}c^{G}_{l}$, ensuring the architecture search does not sacrifice accuracy solely for latency optimization.
If $K = 0$, \sys{} operates without the sub-network, resulting in $g(a) = \sum_{l=1}^{L}c^{G}_{l}$.

\subsection{TEE-aware Search Space}
\label{subsec:search_space}

\subsubsection{Block configuration.}
For P3, we construct a unified search space accommodating diverse backbone architectures, including CNNs and Transformers.
Inspired by lightweight architectures~\cite{howard2017mobilenets, tolstikhin2021mlp}, \sys{} supports both spatial and channel mixing blocks.
As depicted in Figure~\ref{fig:config}, this encompasses token and channel mixing for Transformers, alongside depth-wise and point-wise convolutions for CNNs, enabling the sub-network to optimally balance accuracy and latency.

To minimize latency within the resource-constrained TEE, \sys{} encapsulates these mixing blocks within a bottleneck structure.
Specifically, it employs average pooling for down-projection and a linear layer ($W^{u}$) for up-sampling.
The bottleneck reduces spatial dimensions (e.g., tokens or resolution) for spatial mixing blocks, and down-samples/up-samples channel dimensions for channel mixing blocks.
This design guarantees execution feasibility while preserving representational capacity.
Detailed configurations are provided in Section A of the supplementary material.

\subsubsection{Search factors.}
To systematically formulate the architecture search for the TEE sub-network, we define the configuration $a \in \mathcal{A}$.
Specifically, the configuration $a$ is uniquely determined by two global up-sampling factors ($S^{u}, C^{u}$) and five block-specific factors ($T_{k}, S^{h}_{k}, C^{h}_{k}, S^{d}_{k}, C^{d}_{k}$) for the $k$-th sub-network block.
First, $S^{u}$ and $C^{u}$ represent the global spatial and channel up-sampling dimensions, respectively.
These are maintained as constant values across all blocks to ensure dimensional compatibility when aggregating the outputs of the sub-network blocks.
$T_{k}$ denotes the operation type of the $k$-th block.
For instance, $T_{k} = 0$ indicates an inactive block, whereas $T_{k} = 1$ and $T_{k} = 2$ represent active spatial mixing and channel mixing blocks, respectively.
When the $k$-th block is active, $S^{d}_{k}$ and $C^{d}_{k}$ define its down-projection dimensions for the bottleneck structure, while $S^{h}_{k}$ and $C^{h}_{k}$ denote its internal hidden dimensions.
The detailed value ranges for each search factor are provided in Section A of the supplementary material.

\subsection{TEE-aware Neural Architecture Search}
\label{subsec:nas}

Even with constrained search factors, exploring the entire search space is computationally prohibitive.
For instance, a 13-block Vision Transformer yields over $10^{20}$ possible configurations.
To navigate this vast space and realize P3 efficiently, \sys{} employs Bayesian Optimization (BO) with Gaussian Processes (GP)~\cite{snoek2012practical}.
Specifically, we utilize a sparse axis-aligned subspace GP (SAAS-GP)~\cite{eriksson2021saas} to effectively handle high-dimensional discrete spaces by identifying sparse relevant dimensions.
To reduce computational overhead, \sys{} trains multiple candidate sub-networks in parallel by freezing the backbone weights.
Furthermore, \sys{} minimizes search cost by evaluating the objective $f(\cdot)$ after a single-epoch training on either a subset or the full dataset.

Driven by the NEHVI acquisition function~\cite{daulton2021differentiable}, \sys{} iteratively samples candidate batches to update the SAAS-GP surrogate model.
Upon completion, \sys{} extracts the Pareto-optimal solutions and returns the optimal configuration $a^*$ that maximizes the score $S(a)$ in Eq.~\ref{eq:score}.
A detailed algorithmic procedure is provided in Section B of the supplementary material.

\subsection{Self-Poisoning Learning}
\label{subsec:poisoning}

After identifying the optimal sub-network $a^*$, we train the combined architecture.
Since the TEE's strict architectural constraints inherently limit $a^*$'s representational capacity and degrade accuracy, we propose a novel self-poisoning learning to overcome this without violating P1.
As analyzed in Section~\ref{sec:motivation}, adversarially trained weights actively hinder an attacker's ability to steal sensitive weights.
Leveraging this, our joint objective deliberately "poisons" the standalone backbone, forcing the sub-network to learn highly specialized, complementary representations to maximize the combined accuracy:

\begin{equation}
    \mathcal{L}_{total} = \beta \cdot \mathcal{L}_{CE}(M_{c}(x), y) + (1 - \beta) \cdot \mathcal{L}_{KD}(M_{c}(x), M_{t}(x)) - \lambda \cdot \mathcal{L}_{CE}(M_{b}(x), y),
    \label{eq:poisoning}
\end{equation}
where $M_{b}(x)$, $M_{c}(x)$, and $M_{t}(x)$ are the logits from the standalone backbone, the combined model, and a teacher model, respectively.
The hyperparameters $\beta$ and $\lambda$ balance the task-specific objectives ($\mathcal{L}_{CE}$ and $\mathcal{L}_{KD}$~\cite{hinton2015distilling}) against the adversarial penalty ($-\mathcal{L}_{CE}$) applied to the backbone.
Finally, \sys{} employs gradient clipping to maintain stability against the diverging gradients of the adversarial term.

\section{Experiments}
\label{sec:exp}

\subsection{Experimental Setup}
\label{subsec:setup}

\subsubsection{Experimental environment.}
Our experiments are conducted on NVIDIA RTX A6000 GPUs.
For real-world latency measurements, we deploy an NVIDIA Jetson Orin~\cite{Orin} configured to the MAXN power mode.
It features a 12-core Arm Cortex-A78AE CPU and 48MB of secure memory, utilizing OP-TEE based on Arm TrustZone~\cite{arm_trustzone}.
Following~\cite{sun2025tensorshield, mo2020darknetz}, the TEE sub-network is established using Darknet~\cite{darknet13}, while the REE backbone is GPU-accelerated via ONNX Runtime~\cite{onnxruntime}.
To maximize parallel efficiency, the REE and TEE execute inference asynchronously, exchanging intermediate features exclusively through a shared memory interface.

\subsubsection{Models and datasets.} 
We evaluate the performance of \sys{} on image classification tasks.
Specifically, we employ representative CNN architectures (VGG16-BN~\cite{simonyan2015very} and ResNet-18~\cite{he2016deep}) alongside a Vision Transformer (ViT-Base~\cite{dosovitskiy2021image}) to demonstrate the generalizability of our approach across different architectures.
These models are evaluated on the CIFAR-10, CIFAR-100~\cite{krizhevsky2009learning}, and TinyImageNet~\cite{le2015tiny} datasets.
Following previous works~\cite{li2022teeslice}, CNNs process inputs at $32 \times 32$ for CIFAR and $64 \times 64$ for TinyImageNet, while ViT-Base resizes all images to $224 \times 224$.

\subsubsection{Evaluation metrics.} 
We evaluate \sys{} to validate four design goals: \textbf{security}, \textbf{efficiency}, and \textbf{feasibility}, \textbf{generality}. 
As \sys{} inherently demonstrates generality across diverse architectures, we quantitatively assess the remaining three. 
Security is measured by the surrogate model's accuracy (lower indicates stronger protection). 
Efficiency is evaluated via the victim model's accuracy and end-to-end inference latency. 
Lastly, feasibility is validated by execution performance under real-world hardware constraints.

\subsubsection{Attack configuration.}
Following prior TSDP studies~\cite{sun2025tensorshield, li2022teeslice}, we adopt the threat model of Knockoff Nets~\cite{orekondy2019knockoff}.
Specifically, the attacker initializes a surrogate model using pre-trained weights from public sources, combined with exposed private weights in the victim's REE.
The attacker then queries the victim model using merely 1\% of the training data to collect input-output label pairs for surrogate training.

\subsubsection{Baseline defense approaches.} 
To rigorously evaluate \sys{}, we compare its performance against SOTA TSDP methods from both TBP and PBT paradigms.
For the TBP paradigm, we evaluate DarkneTZ~\cite{mo2020darknetz} and Serdab~\cite{mo2021serdab} (protecting the last and first three layers (or blocks) in the TEE, respectively), Magnitude~\cite{chen2022magnitude} (securing the top 1\% largest weights), and GroupCover~\cite{kim2023groupcover} (obfuscating linear layers in the REE while executing non-linear layers in the TEE).
For the PBT paradigm, TEESlice~\cite{li2022teeslice} secures a sub-network alongside non-linear layers in the TEE, whereas TB-Net~\cite{xiao2023tbnet} protects only the sub-network.~\footnote{Since the official implementations for TEESlice (ViT-Base) and TB-Net are not publicly available, we carefully re-implemented these methods based on the details provided in the original papers.}
Finally, we define two reference scenarios: No-shield (all weights exposed to the REE, enabling direct copying without additional training) and Black-box (no weights leaked, restricting the attacker to pre-trained weights and input-output label queries).

\begin{table*}[t]
\caption{The accuracy of the surrogate model regarding SOTA defense methods. \highlight{Green} highlights lower accuracy of surrogate model than Black-box attack.}
\label{tab:ms_exp}
\centering
\resizebox{1\columnwidth}{!}{%
\begin{tabular}{c|c|ccccccccc}
\hline
Model                         & Dataset    & No-Shield    & DarkneTZ     & Serdab       & Magnitude    & GroupCover          & TEESlice          & TB-Net            & Black-box     & \sys{}             \\ \hline
\multirow{3}{*}{VGG16-BN}     & C10        &  93.70       & 93.07        & 91.49        & 90.51        & \highlight{25.91}   & 67.76             & \highlight{31.66} & 67.37        &  \highlight{14.21} \\
                              & C100       &  74.32       & 62.54        & 72.60        & 69.26        & \highlight{4.75}    & \highlight{21.74} & \highlight{6.82}  & 21.99        &  \highlight{3.12}  \\
                              & T200       &  62.14       & 37.06        & 62.07        & 60.35        & \highlight{2.04}    & \highlight{25.14} & \highlight{3.73}  & 25.52        &  \highlight{0.71}  \\ \hline
\multirow{3}{*}{ResNet-18}    & C10        &  95.44       & 89.93        & 90.86        & 83.01        & \highlight{34.44}   & 57.92             & \highlight{40.18} & 57.56        &  \highlight{12.36} \\
                              & C100       &  79.55       & 40.70        & 75.60        & 70.92        & \highlight{7.30}    & \highlight{21.01} & \highlight{6.41}  & 21.33        &  \highlight{2.20}  \\
                              & T200       &  61.28       & 27.78        & 58.33        & 54.05        & \highlight{3.30}    & 15.51             & \highlight{2.01}  & 14.42        &  \highlight{1.55}  \\ \hline
\multirow{3}{*}{ViT-Base}     & C10        &  98.63       & 97.56        & 97.95        & 98.44        & \highlight{10.00}   & 93.98             & -                 & 93.27        &  \highlight{10.00} \\
                              & C100       &  92.55       & 76.60        & 91.43        & 91.56        & \highlight{1.00}    & 32.11             & -                 & 32.03        &  \highlight{1.41}  \\
                              & T200       &  90.88       & 79.08        & 86.10        & 88.14        & \highlight{0.50}    & 45.23             & -                 & 43.65        &  \highlight{0.53}  \\ \hline
\multicolumn{2}{c|}{Average}               & $1.70\times$ & $1.45\times$ & $1.66\times$ & $1.63\times$ & $0.37\times$        & $1.01\times$      & $0.44\times$      & $1.00\times$ & $0.08\times$       \\ \hline
\end{tabular}
}
\end{table*}

\subsection{Results: Security}

For the architecture search, we set $\alpha=0.5$ and the TEE memory limit to $\mathcal{H}_{limit}=24$MB.
For self-poisoning learning, the adversarial factor $\lambda$ is set to 0.01, 0.001, and 0.005 for VGG16-BN, ResNet-18, and ViT-Base, with gradient clipping thresholds of 0.5, 2.0, and 2.0, respectively.

Table~\ref{tab:ms_exp} compares the attacker's surrogate model accuracy across defense methods.
Most TBP methods fail to prevent model extraction, yielding higher surrogate accuracies than the Black-box baseline, except GroupCover which obfuscates REE weights.
Within the PBT paradigm, TEESlice provides Black-box-level defense, while TB-Net further reduces surrogate accuracy.
Notably, \sys{} consistently records the lowest surrogate accuracy across most models and datasets.
For instance, with ResNet-18 on CIFAR-10, \sys{} drastically suppresses surrogate accuracy to 12.36\%, outperforming the most secure SOTA baseline (GroupCover, 34.44\%).
This demonstrates that adversarially spoiling the backbone network effectively neutralizes model stealing attacks.

\begin{table*}[t]
\caption{Accuracy comparison between \sys{} and fully fine-tuning victim model.}
\label{tab:acc}
\centering
\resizebox{0.85\columnwidth}{!}{
\newcolumntype{C}{>{\centering\arraybackslash}X}
\begin{tabularx}{\columnwidth}{c|CCC|CCC|CCC}
\hline
\multirow{2}{*}{Method} & \multicolumn{3}{c|}{VGG16-BN} & \multicolumn{3}{c|}{ResNet-18} & \multicolumn{3}{c}{ViT-Base}  \\
                        & C10     & C100     & T200     & C10      & C100     & T200     & C10     & C100     & T200     \\ \hline
\sys{}                  & 93.34   & 72.05    & 66.26    & 95.84    & 80.68    & 68.68    & 98.74   & 90.26    & 90.65    \\
Baseline                & 93.70   & 74.32    & 62.14    & 95.44    & 79.55    & 61.28    & 98.63   & 92.55    & 90.88    \\ \hline
\end{tabularx}
}
\end{table*}

\subsection{Results: Efficiency}

\subsubsection{Accuracy.}
Table~\ref{tab:acc} compares the classification accuracy of the combined model (the backbone coupled with the sub-network) searched and trained via \sys{} against the fully fine-tuning victim model (No-Shield).
\sys{} achieves comparable or even superior accuracy compared to the baseline across most models and datasets.
Remarkably, for ResNet-18 on TinyImageNet, \sys{} demonstrates a significant accuracy improvement of approximately 7.4\%p (68.68\% vs. 61.28\%).
This highlights that \sys{} does not sacrifice accuracy to enforce strict security; rather, \sys{} often enhances the model's inherent performance.

\subsubsection{Latency.}
\begin{figure}[t]
    \centering
    \includegraphics[width=\linewidth]{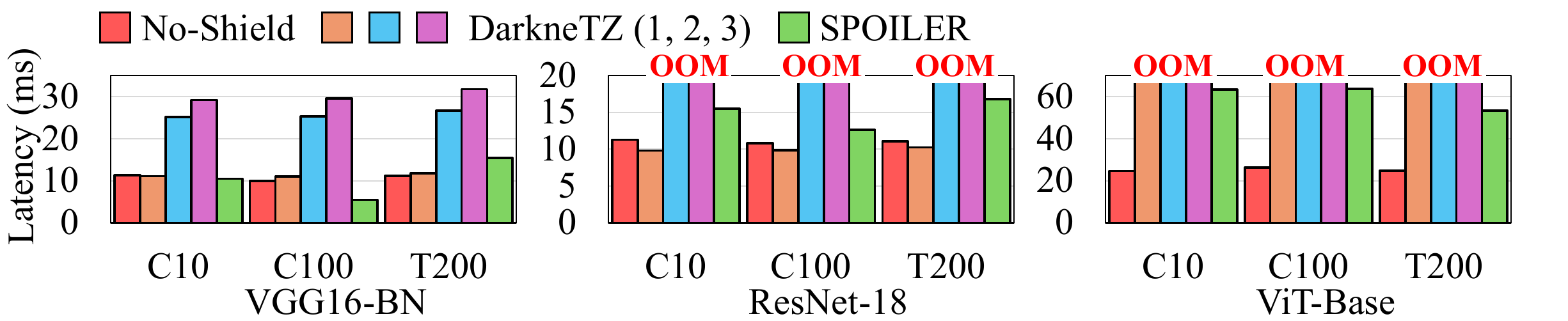}
    \caption{Inference latency comparison of No-shield, DarkneTZ, and \sys{}.}
    \label{fig:execution_time}
\end{figure}

To demonstrate the efficiency of \sys{}, we average 100 latency measurements on the NVIDIA Jetson Orin.
As shown in Figure~\ref{fig:execution_time}, we compare \sys{} against an unprotected GPU baseline (No-shield) and DarkneTZ (isolating the final 1, 2, and 3 layers).
DarkneTZ inherently suffers from the critical flaws shared by all sequential TSDP methods: strict sequential bottlenecks and hardware-oblivious memory allocation.
Consequently, isolating just two layers often triggers Out-Of-Memory (OOM) faults, while layer-by-layer weight loading incurs high I/O overheads.
Conversely, by leveraging NAS and strictly bounding the secure memory via $\mathcal{H}_{limit}$, \sys{} easily fits and executes the sub-networks of heavy models like ViT-Base in a single pass.
By enabling asynchronous parallel execution, \sys{} significantly outperforms TSDP baselines.
Notably, for VGG16-BN on CIFAR-100, \sys{} even surpasses No-shield's speed, as the independent sub-network can terminate without waiting for the entire backbone inference to finish.

\begin{figure}[t]
    \centering
    \includegraphics[width=\linewidth]{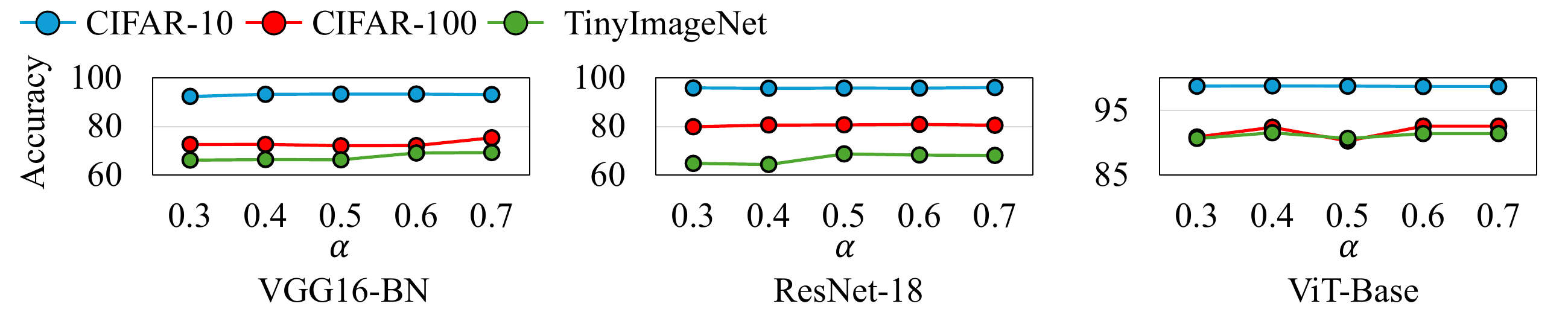}
    \caption{Accuracy of \sys{} across varying $\alpha$.}
    \label{fig:alpha}
\end{figure}

\subsubsection{Trade-off between accuracy and latency.}

Figure~\ref{fig:alpha} presents the accuracy of sub-network configurations derived across varying values of $\alpha$ in Eq.~\eqref{eq:score}. 
As $\alpha$ increases, the task accuracy remains stable, exhibiting a marginal upward trend.
This stability demonstrates the efficacy of \sys{}'s architectural design, indicating that \sys{} consistently isolates high-capacity sub-networks regardless of minor hyperparameter shifts.
Crucially, the accuracy benefits identified during the NAS phase are seamlessly preserved in the final models, even after self-poisoning learning.
This end-to-end stability allows \sys{} to offer multiple valid configuration candidates from just a single search phase, ensuring high accuracy for diverse user constraints.

\begin{figure}[t]
    \centering
    \includegraphics[width=\linewidth]{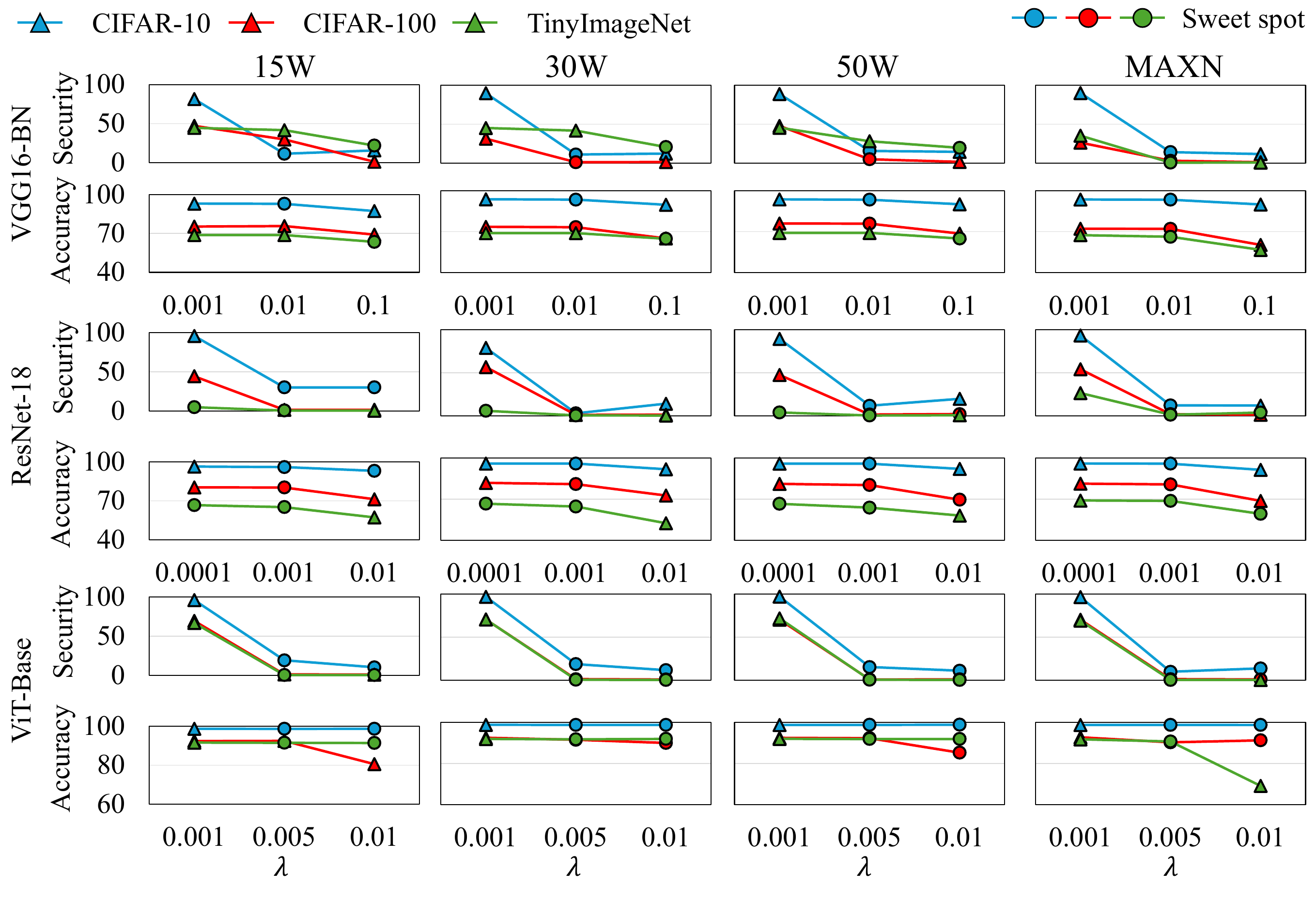}
    \caption{Performance of \sys{} under varying power modes (15W, 30W, 50W, and MAXN) on the Jetson Orin.}
    \label{fig:scalability}
\end{figure}

\subsection{Results: Feasibility}

To simulate diverse edge devices, we evaluate \sys{} across four Jetson Orin power modes (15W, 30W, 50W, and MAXN).
Figure~\ref{fig:scalability} illustrates the security-accuracy trade-off controlled by the adversarial factor $\lambda$ ($\alpha, \beta=0.5$).
\sys{} utilizes $\lambda$ as a flexible control knob between security and accuracy: a higher $\lambda$ enforces stricter security against model stealing, whereas a lower $\lambda$ prioritizes the primary task accuracy.
Notably, the circle markers in the figure highlight the optimal configurations (the `sweet spots') where \sys{} successfully suppresses the surrogate accuracy below the Black-box baseline while restricting the accuracy drop to within a mere 3\%p compared to the fully fine-tuned victim model.
Crucially, at least one such optimal configuration exists across all tested power modes, datasets, and architectures.
This demonstrates that by calibrating $\lambda$, \sys{} seamlessly adapts to various hardware constraints, thoroughly spoiling backbone weights to prevent extraction without sacrificing inherent performance.

\section{Related Work}
\label{sec:related_work}

\subsection{TEE-Shielded Partitioning}
Existing TSDP methods fall into two categories: obfuscation-based and static partitioning.
Obfuscation methods (e.g., ShadowNet~\cite{sun2023shadownet}, TransLinkGuard~\cite{li2024translinkguard}) offload transformed layers to GPUs but incur prohibitive cryptographic overheads during the encryption and decryption phases~\cite{kim2023groupcover}.
Conversely, static partitioning frameworks (e.g., DarkneTZ~\cite{mo2020darknetz}, TBNet~\cite{xiao2023tbnet}, TEESlice~\cite{li2022teeslice}) isolate sensitive layers within the TEE.
However, their reliance on strict sequential dependencies forces highly inefficient lock-step synchronization between the TEE (CPU) and REE (GPU).
This architectural rigidity causes severe resource idling and exacerbates frequent communication overhead.
To overcome these limitations, \sys{} employs TEE-aware NAS to dynamically decouple the TEE sub-network from rigid backbone constraints.

\subsection{NAS for Efficient Inference}
Hardware-aware NAS frameworks (e.g., MnasNet~\cite{tan2019mnasnet}, ProxylessNAS~\cite{cai2018proxylessnas}, NetAdapt~\cite{yang2018netadapt}, MCUNet~\cite{lin2020mcunet}) co-optimize architectures for task utility and device constraints like latency or memory.
Recent advances also explore supernet training (OFA~\cite{cai2019once}) and adversarial robustness (RobNets~\cite{guo2020meet}).
Furthermore, paradigms like AutoPEFT~\cite{zhou2024autopeft} have shifted focus toward searching for lightweight sub-modules, such as adapters, rather than optimizing the entire backbone.
However, whether targeting full architectures or sub-modules, existing NAS literature predominantly focuses on accuracy or input robustness.
They do not explicitly account for model confidentiality and the unique constraints of TEEs, such as stringent secure memory limits and severe cross-domain synchronization overheads.
\sys{} bridges this gap by explicitly integrating TEE-specific hardware constraints into the search space, simultaneously optimizing for inference efficiency and logical isolation.

\section{Conclusion}
\label{sec:conclusion}

In this paper, we introduce \sys{}, a novel defense framework designed to protect model privacy against model stealing attacks on resource-constrained edge devices.
To overcome the limitations of existing TSDP paradigms bound by fixed topologies and costly obfuscation, \sys{} pioneers the SBT paradigm.
Specifically, it leverages TEE-aware NAS to decouple lightweight TEE sub-networks for parallel execution.
Furthermore, to actively disrupt model extraction while preserving the primary task utility, we propose an adversarial self-poisoning learning.
Extensive hardware evaluations confirm \sys{}'s seamless adaptability to stringent constraints, jointly optimizing security, latency, and accuracy.

\clearpage  


%
%
\bibliographystyle{splncs04}
\bibliography{main}

@String(CVPR  = {IEEE Conf. Comput. Vis. Pattern Recog.})

@String(ECCV  = {Eur. Conf. Comput. Vis.})

@String(NeurIPS = {Adv. Neural Inform. Process. Syst.})

@String(ICML  = {Int. Conf. Mach. Learn.})

@String(ICLR  = {Int. Conf. Learn. Represent.})

@String(CVPR  = {CVPR})

@String(ECCV  = {ECCV})

@String(NeurIPS = {NeurIPS})

@String(ICML  = {ICML})

@String(ICLR  = {ICLR})

@article{lecun2015deep,
  title={Deep learning},
  author={LeCun, Yann and Bengio, Yoshua and Hinton, Geoffrey},
  journal={nature},
  volume={521},
  number={7553},
  pages={436--444},
  year={2015}
}

@inproceedings{tramer2016stealing,
  title={Stealing Machine Learning Models via Prediction {APIs}},
  author={Tram{\`e}r, Florian and Zhang, Fan and Juels, Ari and Reiter, Michael K and Ristenpart, Thomas},
  booktitle={25th USENIX security symposium (USENIX Security 16)},
  year={2016},
  pages={601--618}
}

@inproceedings{orekondy2019knockoff,
  title={Knockoff nets: Stealing functionality of black-box models},
  author={Orekondy, Tribhuvanesh and Schiele, Bernt and Fritz, Mario},
  booktitle={Proceedings of the IEEE/CVF Conference on Computer Vision and Pattern Recognition (CVPR)},
  pages={4954--4963},
  year={2019}
}

@inproceedings{juuti2019prada,
  title={{PRADA}: Protecting against {DNN} Model Stealing Attacks},
  author={Juuti, Mika and Szyller, Sebastian and Marchal, Samuel and Asokan, N},
  booktitle={2019 IEEE European Symposium on Security and Privacy (EuroS\&P)},
  pages={512--527},
  year={2019},
}

@manual{arm_trustzone,
  title={{ARM} Security Technology: Building a Secure System using {TrustZone} Technology},
  author={{ARM Limited}},
  year={2009},
  note={White paper}
}

@online{smc_convention,
  title={{SMC} calling convention {[Online]}},
  author={{ARM Limited}},
  howpublished = {{Available: }\url{https://developer.arm.com/documentation/den0028/h/}},
  year={2026}
}

@article{costan2016intel,
  title={Intel {SGX} explained},
  author={Costan, Victor and Devadas, Srinivas},
  journal={Cryptology ePrint Archive},
  year={2016}
}

@inproceedings{mo2020darknetz,
  title={{DarkneTZ}: Towards Model Privacy at the Edge using Trusted Execution Environments},
  author={Mo, Fan and Shamsabadi, Ali Shahin and Katevas, Kleomenis and Demetriou, Soteris and Leontiadis, Ilias and Cavallaro, Andrea and Haddadi, Hamed},
  booktitle={Proceedings of the 18th International Conference on Mobile Systems, Applications, and Services (MobiSys)},
  pages={161--174},
  year={2020}
}

@inproceedings{mo2021serdab,
  title={Serdab: An {IoT} framework for partitioning neural networks computation across multiple enclaves},
  author={Elgamal, Tarek and Nahrstedt, Klara},
  booktitle={2020 20th IEEE/ACM International Symposium on Cluster, Cloud and Internet Computing (CCGRID)},
  pages={519--528},
  year={2020}
}

@inproceedings{tramer2018slalom,
  title={Slalom: Fast, verifiable and private execution of neural networks in trusted hardware},
  author={Tram{\`e}r, Florian and Boneh, Dan},
  booktitle={International Conference on Learning Representations (ICLR)},
  year={2019}
}

@inproceedings{lin2020chaotic,
  title={Chaotic weights: A novel approach to protect intellectual property of deep neural networks},
  author={Lin, Ning and Chen, Xiaoming and Lu, Hang and Li, Xiaowei},
  booktitle={IEEE Transactions on Computer-Aided Design of Integrated Circuits and Systems (TCAD)},
  year={2020},
  volume={40},
  pages={1327--1339},
}

@inproceedings{xiao2023tbnet,
  title={{TBNet}: a neural architectural defense framework facilitating DNN model protection in trusted execution environments},
  author={Liu, Ziyu and Zhou, Tong and Luo, Yukui and Xu, Xiaolin},
  booktitle={Proceedings of the 61st ACM/IEEE Design Automation Conference (DAC)},
  pages={1--6},
  year={2024}
}

@inproceedings{zoph2016neural,
  title={Neural Architecture Search with Reinforcement Learning},
  author={Zoph, Barret and Le, Quoc V},
  booktitle={International Conference on Learning Representations (ICLR)},
  year={2017}
}

@inproceedings{cai2018proxylessnas,
  title={{ProxylessNAS}: Direct Neural Architecture Search on Target Task and Hardware},
  author={Cai, Han and Zhu, Ligeng and Han, Song},
  booktitle={International Conference on Learning Representations (ICLR)},
  year={2019}
}

@article{sabt2015trusted,
  title={Trusted execution environment: What it is, and what it is not},
  author={Sabt, Mohamed and Achemlal, Mohammed and Bouabdallah, Abdelmadjid},
  journal={IEEE Trustcom/BigDataSE/ISPA},
  volume={1},
  pages={57--64},
  year={2015}
}

@article{pinto2019demystifying,
  title={Demystifying Arm TrustZone: A comprehensive survey},
  author={Pinto, Sandro and Santos, Nuno},
  journal={ACM Computing Surveys (CSUR)},
  volume={51},
  number={6},
  pages={1--36},
  year={2019}
}

@inproceedings{jagielski2020high,
  title={High accuracy and high fidelity extraction of neural networks},
  author={Jagielski, Matthew and Carlini, Nicholas and Berthelot, David and Kurakin, Alex and Papernot, Nicolas},
  booktitle={29th USENIX security symposium (USENIX Security 20)},
  pages={1345--1362},
  year={2020}
}

@article{chen2022magnitude,
  title={Model protection: {Real-time} privacy-preserving inference service for model privacy at the edge},
  author={Hou, Jiahui and Liu, Huiqi and Liu, Yunxin and Wang, Yu and Wan, Peng-Jun and Li, Xiang-Yang},
  journal={IEEE Transactions on Dependable and Secure Computing (TDSC)},
  volume={19},
  number={6},
  pages={4270--4284},
  year={2021}
}

@inproceedings{kim2023groupcover,
  title={Groupcover: A secure, efficient and scalable inference framework for on-device model protection based on tees},
  author={Zhang, Zheng and Wang, Na and Zhang, Ziqi and Zhang, Yao and Zhang, Tianyi and Liu, Jianwei and Wu, Ye},
  booktitle={Forty-first international conference on machine learning (ICML)},
  year={2024}
}

@inproceedings{li2022teeslice,
  title={{TeeSlice}: slicing {DNN} models for secure and efficient deployment},
  author={Zhang, Ziqi and Ng, Lucien KL and Liu, Bingyan and Cai, Yifeng and Li, Ding and Guo, Yao and Chen, Xiangqun},
  booktitle={Proceedings of the 2nd ACM International Workshop on AI and Software Testing/Analysis (AISTA)},
  year={2022},
  pages={1--8}
}

@article{howard2017mobilenets,
  title={{MobileNets}: Efficient Convolutional Neural Networks for Mobile Vision Applications},
  author={Howard, Andrew G and Zhu, Menglong and Chen, Bo and Kalenichenko, Dmitry and Wang, Weijun and Weyand, Tobias and Andreetto, Marco and Adam, Hartwig},
  journal={arXiv preprint arXiv:1704.04861},
  year={2017}
}

@article{tolstikhin2021mlp,
  title={Mlp-mixer: An all-mlp architecture for vision},
  author={Tolstikhin, Ilya O and Houlsby, Neil and Kolesnikov, Alexander and Beyer, Lucas and Zhai, Xiaohua and Unterthiner, Thomas and Yung, Jessica and Steiner, Andreas and Keysers, Daniel and Uszkoreit, Jakob and others},
  journal={Advances in neural information processing systems},
  volume={34},
  pages={24261--24272},
  year={2021}
}

@inproceedings{snoek2012practical,
  title={Practical {Bayesian} optimization of machine learning algorithms},
  author={Snoek, Jasper and Larochelle, Hugo and Adams, Ryan P},
  booktitle={Advances in Neural Information Processing Systems (NeurIPS)},
  year={2012},
  volume={25}
}

@inproceedings{eriksson2021saas,
  title={High-dimensional {Bayesian} optimization with sparse axis-aligned subspaces},
  author={Eriksson, David and Jankowiak, Martin},
  booktitle={Uncertainty in Artificial Intelligence (UAI)},
  pages={493--503},
  year={2021}
}

@inproceedings{daulton2021differentiable,
  title={Differentiable Expected Hypervolume Improvement for Parallel Multi-Objective {Bayesian} Optimization},
  author={Daulton, Samuel and Balandat, Maximilian and Bakshy, Eytan},
  booktitle={Advances in Neural Information Processing Systems (NeurIPS)},
  volume={33},
  pages={9851--9864},
  year={2020}
}

@inproceedings{sun2023shadownet,
  title={{ShadowNet}: A Secure and Efficient On-device Model Inference System for Convolutional Neural Networks},
  author={Sun, Zhichuang and Sun, Ruimin and Liu, Changming and Chowdhury, Amrita Roy and Lu, Long and Jha, Somesh},
  booktitle={2023 IEEE Symposium on Security and Privacy (SP)},
  pages={1596--1612},
  year={2023}
}

@inproceedings{li2024translinkguard,
  title={{TransLinkGuard}: Safeguarding Transformer Models Against Model Stealing in Edge Deployment},
  author={Li, Qinfeng and Shen, Zhiqiang and Qin, Zhenghan and Xie, Yangfan and Zhang, Xuhong and Du, Tianyu and Cheng, Sheng and Wang, Xun and Yin, Jianwei},
  booktitle={Proceedings of the 32nd ACM International Conference on Multimedia (MM)},
  year={2024},
  pages={3479--3488}
}

@inproceedings{sun2025tensorshield,
  title={{TensorShield}: Safeguarding On-Device Inference by Shielding Critical {DNN} Tensors with {TEE}},
  author={Sun, Tong and Jiang, Bowen and Lin, Hailong and Li, Borui and Teng, Yixiao and Gao, Yi and Dong, Wei},
  booktitle={Proceedings of the 2025 ACM SIGSAC Conference on Computer and Communications Security (CCS)},
  year={2025}
}

@inproceedings{lin2020mcunet,
  title={{MCUNet}: Tiny Deep Learning on {IoT} Devices},
  author={Lin, Ji and Chen, Wei-Ming and Lin, Yujun and Gan, Chuang and Han, Song},
  booktitle={Advances in Neural Information Processing Systems (NeurIPS)},
  year={2020},
  volume={33},
  pages={11711--11722}
}

@inproceedings{tan2019mnasnet,
  title={{MnasNet}: Platform-Aware Neural Architecture Search for Mobile},
  author={Tan, Mingxing and Chen, Bo and Pang, Ruoming and Vasudevan, Vijay and Sandler, Mark and Howard, Andrew and Le, Quoc V},
  booktitle={Proceedings of the IEEE/CVF Conference on Computer Vision and Pattern Recognition (CVPR)},
  pages={2820--2828},
  year={2019}
}

@inproceedings{cai2019once,
  title={{Once-for-All}: Train One Network and Specialize it for Efficient Deployment},
  author={Cai, Han and Gan, Chuang and Wang, Tianzhe and Zhang, Zhekai and Han, Song},
  booktitle={International Conference on Learning Representations (ICLR)},
  year={2020}
}

@inproceedings{guo2020meet,
  title={When {NAS} Meets Robustness: In Search of Robust Architectures against Adversarial Attacks},
  author={Guo, Minghao and Yang, Yuzhe and Xu, Rui and Liu, Ziwei and Lin, Dahua},
  booktitle={Proceedings of the IEEE/CVF Conference on Computer Vision and Pattern Recognition (CVPR)},
  pages={631--640},
  year={2020}
}

@inproceedings{yang2018netadapt,
  title={{NetAdapt}: Platform-Aware Neural Network Adaptation for Mobile Applications},
  author={Yang, Tien-Ju and Howard, Andrew and Chen, Bo and Zhang, Xiao and Go, Alec and Sandler, Mark and Sze, Vivienne and Adam, Hartwig},
  booktitle={Proceedings of the European conference on computer vision (ECCV)},
  pages={285--300},
  year={2018}
}

@online{Orin,
  title = {{NVIDIA Orin Developer Kit. [Online]}},
  howpublished = {{Available: }\url{https://www.nvidia.com/ko-kr/autonomous-machines/embedded-systems/jetson-orin/}}
}

@article{hinton2015distilling,
  title={Distilling the knowledge in a neural network},
  author={Hinton, Geoffrey and Vinyals, Oriol and Dean, Jeff},
  journal={arXiv preprint arXiv:1503.02531},
  year={2015}
}

@misc{darknet13,
  author={Joseph Redmon},
  title={Darknet: Open Source Neural Networks in {C}},
  howpublished={\url{https://pjreddie.com/darknet/}},
  year={2013--2016}
}

@misc{onnxruntime,
  title={{ONNX Runtime}},
  author={{ONNX Runtime developers}},
  year={2021},
  howpublished={\url{https://onnxruntime.ai/}}
}

@inproceedings{simonyan2015very,
  title={Very deep convolutional networks for large-scale image recognition},
  author={Simonyan, Karen and Zisserman, Andrew},
  booktitle={International Conference on Learning Representations (ICLR)},
  year={2015}
}

@inproceedings{he2016deep,
  title={Deep residual learning for image recognition},
  author={He, Kaiming and Zhang, Xiangyu and Ren, Shaoqing and Sun, Jian},
  booktitle={Proceedings of the IEEE/CVF Conference on Computer Vision and Pattern Recognition (CVPR)},
  pages={770--778},
  year={2016}
}

@inproceedings{dosovitskiy2021image,
  title={An image is worth 16x16 words: {Transformers} for image recognition at scale},
  author={Dosovitskiy, Alexey and Beyer, Lucas and Kolesnikov, Alexander and Weissenborn, Dirk and Zhai, Xiaohua and Unterthiner, Thomas and Dehghani, Mostafa and Minderer, Matthias and Heigold, Georg and Gelly, Sylvain and others},
  booktitle={International Conference on Learning Representations (ICLR)},
  year={2021}
}

@article{krizhevsky2009learning,
  title={Learning multiple layers of features from tiny images},
  author={Krizhevsky, Alex and Hinton, Geoffrey and others},
  year={2009},
  publisher={Toronto, ON, Canada}
}

@article{le2015tiny,
  title={{Tiny Imagenet} visual recognition challenge},
  author={Le, Yann and Yang, Xuan and others},
  journal={CS 231N},
  volume={7},
  number={7},
  pages={3},
  year={2015}
}

@article{zhou2024autopeft,
  title={Autopeft: Automatic configuration search for parameter-efficient fine-tuning},
  author={Zhou, Han and Wan, Xingchen and Vuli{\'c}, Ivan and Korhonen, Anna},
  journal={Transactions of the Association for Computational Linguistics},
  volume={12},
  pages={525--542},
  year={2024},
  publisher={MIT Press One Broadway, 12th Floor, Cambridge, Massachusetts 02142, USA~…}
}
\end{document}